\documentclass[aps,pre,letterpaper,twocolumn,floatfix,showpacs]{revtex4}
\usepackage{graphicx,amsmath}
\usepackage{graphicx}
\usepackage{dcolumn}
\usepackage{bm}
\usepackage{url}

\def\I.#1{\it #1}
\def\B.#1{{\bbox#1}}
\def\C.#1{{\cal #1}}

\newcommand\eq{\begin{equation}}
\newcommand\en{\end{equation}}

\begin{document}

\title{Spreading of thin films assisted by thermal fluctuations}
\author{Benny Davidovitch$^1$, Esteban Moro$^2$, and Howard A. Stone$^1$}
\affiliation{
 $^1$ Division of Engineering and Applied Sciences,
 Harvard University,
 Cambridge, MA 02138 \\
 $^2$ Grupo Interdisciplinar de Systemas Complejos (GISC)
 and Departmento de Mathematics, Universidad Carlos III de
 Madrid, Avda. de la Universidad 30, E-28911, Legans, Spain}

\date{\today}
\begin{abstract}
We study the spreading of viscous drops on a solid substrate,
taking into account the effects of thermal fluctuations in the
fluid momentum. A nonlinear stochastic lubrication equation is
derived, and studied using numerical simulations and scaling
analysis. We show that asymptotically spreading drops admit
self-similar shapes, whose average radii can increase at rates
much faster than these predicted by Tanner's law. We discuss the
physical realizability of our results for thin molecular and
complex fluid films, and predict that such phenomenon can in
principal be observed in various flow geometries.
\end{abstract}


\maketitle

Water drops spreading on a table and oil drops lubricating a pan
are two common examples of a phenomenon encountered frequently in
the kitchen as well as in natural and industrial environments:
Spreading of liquids on solid surfaces. Despite its prevalence and
the basic hydrodynamic principles involved, it was not until the
late 1970's that the asymptotic rate of spreading processes was
found by Tanner \cite{Tanner} for surface-tension dominated flows.
The spatial scale $\ell$ of a viscous drop spreading on a smooth
plane increases asymptotically in time as $\ell \sim t^{z}$, where
$z = 1/10,1/7$ for radially symmetric two-dimensional and
one-dimensional flow geometries, respectively. This asymptotic
response has been found in many molecular and polymeric drops,
whose decreasing thickness has been detected down to $100 \ nm$
\cite{McKinley}. As the field of nano-fluidics is evolving towards
formation of thinner and thinner liquid films, theoretical tools
are needed to describe flow patterns in such geometries. However,
the applicability of classical hydrodynamic theory for these
systems is questionable. While the necessity to incorporate van
der Waals (vdW) fluid-solid attraction was recognized long ago
\cite{DeGennes}, other fundamental aspects have never been fully
addressed. In particular: Does a three-dimensional hydrodynamic
description hold for a film whose thickness is just few molecular
layers? What are the effects of thermal fluctuations on the
deterministic hydrodynamics at such small scales?

To resolve these questions, extensive molecular dynamics (MD) or
lattice-gas (LG) simulations of flow in liquid films are required
to allow comparison and quantify deviations from a regular
hydrodynamic theory \cite{Koplik-Freund-He}. While a full
resolution of these questions is still not available within
current computational possibilities, a few recent studies are
indicative. For example, Abraham {\it et} al. \cite{Esteban} used
LG algorithm to study flow in a precursor film associated with a
spreading drop, and demonstrated significant deviations from the
predictions of a hydrodynamic model \cite{DeGennes}. On the other
hand, MD simulations of nanojets, whose initial radius was around
10 molecular diameters, were shown to be qualitatively consistent
with simulations of a stochastic Navier-Stokes (NS) equation,
where the viscous stress tensor was supplemented by a stochastic
tensor whose temperature-dependent magnitude is determined from
the fluctuation-dissipation theorem \cite{Landman}. The emerging
picture is that, at least in some cases, a hydrodynamic
description can still be used as a quantitative tool in studying
flow of nano-fluids, however modifications of the classical
equations are required.

In this Letter we take one further step forward by exploring the
influence of thermal fluctuations on the shape and rate of
spreading of nano-dimension drops, while assuming a generalized NS
equation holds, similarly to \cite{Landman}. We should note,
however, that in addition to thermal fluctuations other
modifications of the hydrodynamic equations that stem from density
variations near the interfaces \cite{Pismen}, might be necessary
in this regime.

Let us start by considering the dynamics of the height $h(x,y,t)$
of an infinite incompressible planar viscous fluid film on top of
a smooth solid surface, located at $h=0$, as depicted in Fig 1. We
consider highly viscous fluids, such that inertia can be
neglected. The mass conserving dynamics of long wave length
fluctuations of the surface, $|\nabla h| \ll 1$, is described by
the lubrication equation
\begin{equation}
\frac{\partial h}{\partial t} = \frac{1}{3\eta} \nabla \cdot (h^3
\nabla p) \ , \label{lubrication1}
\end{equation}
where $\eta$ is the viscosity and $p$ is the pressure. The
derivation of Eq. (\ref{lubrication1}) from the NS equation is a
standard exercise in fluid mechanics \cite{Batchelor}. Spatial
variations of the pressure associated with fluctuations of the
liquid-vapor interface result from several sources: gravity,
surface tension, and VdW attraction with the solid surface:
\begin{equation}
p  = \rho g h - \gamma \nabla^2 h + A/h^3 \ , \label{pressure}
\end{equation}
where $\rho$ is the fluid density, $g$ is the gravitational
acceleration, $\gamma$ is the liquid-vapor surface tension, and
$A$ is the Hamaker constant. Tanner's law corresponds to
similarity solutions of Eq. (\ref{lubrication1})
\begin{equation}
h(\vec{x},t) = |x|^{-\beta} f(|x|/t^{z}) \label{scalingform}
\end{equation}
in the surface-tension-dominated regime $|\gamma \nabla^2 h| \gg
|\rho g h|, |A/h^3| $, where the exponent $\beta=1,2$ is
determined by requiring volume conservation $ V = \int d^{d}
\vec{x} h(\vec{x},t)$, yielding $\beta=1$ for one-dimensional
($d=1$) and $\beta=2$ for two-dimensional ($d=2$) flow geometries,
respectively.

The central equation of this paper is a stochastic generalization
of Eq. (\ref{lubrication1}):
\begin{equation}
\frac{\partial h}{\partial t} = \frac{1}{3\eta} \nabla \cdot (h^3
 \nabla  p)  + \sqrt{\frac{2k_B T}{3 \eta}} \nabla \cdot [h^{3/2}
\xi (\vec{x},t)] \ , \label{lubrication3}
\end{equation}
which captures effects of thermal fluctuations on the surface
dynamics. Here, $\xi (\vec{x},t)$ is a spatio-temporal Gaussian
white noise. Principally, Eq. (\ref{lubrication3}) can be derived
from the full three-dimensional NS equation, similarly to Eq.
(\ref{lubrication1}), by adding a stochastic stress, representing
thermal fluctuations of the fluid momentum, to the viscous stress
tensor \cite{Landau}. We can avoid however such a tedious
derivation by considering a reduced, linear version of Eq.
(\ref{lubrication1}), and use the fluctuation-dissipation theorem
to find the correct magnitude of a Langevin term that gives rise
to equipartition of the thermal energy density carried by its
eigenmodes $h_{\vec{q}} (\vec{x},t) = H + \delta h_{\vec{q}}(t)
\cos (\vec{q} \cdot \vec{x})$. Here $|\delta h_{\vec{q}} /H| \ll
1$, $H$ is the average thickness of the film, and $\vec{q}$ is a
planar wave vector. Namely, the linear eigenmodes of the
stochastic surface dynamics are required to satisfy:
\begin{equation}
\Gamma_{|q|} \frac{\partial \delta h_{\vec{q}}}{\partial t}  +
p_{\vec{q}} = \sqrt{2\Gamma_{|q|} k_B T} \  \xi_{\vec{q}}(t)  \ ,
\label{equipartition}
\end{equation}
where $p_{\vec{q}} = (\rho g + \gamma |q|^2  - 3A/H^4) \delta
h_{\vec{q}} \cos(\vec{q} \cdot \vec{x})$ is the pressure
(mechanical energy density) of a surface eigenmode,
$\Gamma_{\vec{q}} = 3\eta / |q|^2 H^3 $ is its friction
coefficient, and $\xi_{\vec{q}}(t)$ is the spatial Fourier
transform of $\xi (\vec{x},t)$. Notice that the long wavelength
approximation underlying Eq. (\ref{lubrication1}) implies $|q| H
\ll 1$. Dividing both sides of Eq. (\ref{equipartition}) by
$\Gamma_{|q|}$ and taking the inverse Fourier transform we obtain:
\begin{equation}
\frac{\partial \delta h}{\partial t} = \frac{1}{3\eta} \nabla
\cdot (H^3 \nabla p) + \sqrt{\frac{2k_BT H^3}{3\eta}} \ \nabla
\cdot \xi (\vec{x},t) \ . \label{lubrication2}
\end{equation}
The linear Eq. (\ref{lubrication2}) describes near equilibrium
thermal fluctuations of a surface, $|\delta h|\ll H$. The
spreading dynamics of a drop that does not satisfy this condition
must be described by a nonlinear equation. To this end, notice
that Eq. (\ref{lubrication1}) can be recovered from the
deterministic part of Eq. (\ref{lubrication2}) by making the
transformation $H,\delta h \to h$ and requiring the resulting
equation to conserve fluid mass. By following exactly the same
steps the nonlinear Langevin Eq. (\ref{lubrication3}) can be
derived from Eq. (\ref{lubrication2}).

A word of caution is in order here. As for any extrapolation of
the fluctuation-dissipation theorem to nonlinear, far from
equilibrium dynamics, a local equilibrium assumption must be made
\cite{Siggia}. Namely, the description of the surface dynamics
with Eq. (\ref{lubrication3}) assumes that the magnitude of
thermal fluctuations of the liquid-vapor interface at $\vec{x}$ is
determined solely by the local value of the surface height
$h(\vec{x},t)$. This assumption is justified only for thermal
fluctuations whose wavelength $\lambda \ll \lambda^{*} (\vec{x},t)
= h / |\nabla h|$. The relaxation dynamics of fluctuations whose
wavelength $\lambda > \lambda^{*}(\vec{x},t)$ is strongly coupled
to interface fluctuations, and thus their magnitude cannot be
assumed to be given by the near-equilibrium result
(\ref{lubrication3}). Including these modes in the stochastic
analysis involves advanced methods \cite{Siggia}, and will not be
pursued here. Since we expect $\lambda^{*}(\vec{x},t) \to \infty$
as $t \to \infty$, our local equilibrium assumption becomes
asymptotically correct. In addition, the number of linear
eigenmodes with $\lambda > \lambda^{*}$ scales as
$(\lambda^{*}/L)^{d-1}$, where $L$ is the lateral size of the
drop. Therefore, we expect Eq. (\ref{lubrication3}) to provide a
better description of the pre-asymptotic dynamics in
two-dimensional than in one-dimensional geometries.

In studying Eq. (\ref{lubrication3}), our basic motivation was to
understand the possible effects thermal fluctuations may have on
the asymptotic rates of spreading, e.g. by modifying Tanner's law.
With this view, we focused our analysis on two characteristic flow
geometries: (i) one-dimensional drops confined in a channel of
width $W$ (Fig. 1), and (ii) two-dimensional radially symmetric
drops.
\begin{figure}
\begin{center}
\includegraphics[width=2in,clip=]{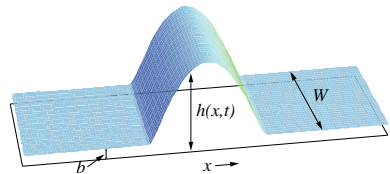}
\caption{\label{fig1} Schematic representation of a
one-dimensional spreading drop, confined in a channel of width
$W$. The precursor layer of height $b$ is also shown.}
\end{center}
\end{figure}
In the first case, Eq. (\ref{lubrication3}) assumes the
form:
\begin{equation}
\frac{\partial h}{\partial t} =  - \frac{\gamma}{3\eta}
\frac{\partial}{\partial x} (h^3  \frac{\partial^3 h}{\partial
x^3}) + \sqrt{\frac{2k_BT}{3\eta W} } \frac{\partial}{\partial x}
\left [ h^{3/2} \xi (x,t) \right ] \ . \label{lubrication31d}
\end{equation}
In the second case, Eq. (\ref{lubrication3}) becomes:
\begin{equation}
\frac{\partial h}{\partial t} =  - \frac{\gamma}{3\eta}
\frac{1}{r}\frac{\partial}{\partial r} \left [ r h^3
\frac{\partial}{\partial r} (\frac{1}{r} \frac{\partial}{\partial
r} r\frac{\partial h}{\partial r} ) \right ] +
\sqrt{\frac{2k_BT}{3\pi \eta}} \frac{1}{r}
\frac{\partial}{\partial r} \left [ h^{3/2} r^{1/2}\xi (r,t)
\right ] \ . \label{lubrication32d}
\end{equation}
We simulated the volume-conserving spreading of a drop dominated
by the dynamics (\ref{lubrication31d}) by using
finite-difference-based computational techniques as in
\cite{numerical}, which guarantee non-negativity of the field
$h(x,t)$ for the deterministic part of Eq. (\ref{lubrication31d}).
In our simulations, the noise term is included in the RHS of a
spatial-temporal discrete version of Eq. (\ref{lubrication31d}),
which is advanced in time through an implicit method.
Non-negativity of $h(x,t)$ is physically implemented by a
short-range repulsive potential between the liquid and the solid
substrate, which in principle should be included in the pressure
in Eq. (\ref{lubrication3}). We avoid the explicit use of such a
potential, by allowing numerical noise realizations only if they
preserve non-negativity of $h(x,t)$. Such a procedure induces
correlations in the otherwise white noise field, which are
unavoidable if the repulsive potential is not introduced
explicitly. In our simulations we use a non-dimensional version of
Eq. (\ref{lubrication31d}):
\begin{equation}
\frac{\partial \tilde{h}}{\partial \tilde{t}} =
 - \frac{\partial}{\partial \tilde{x}} ( \tilde{h}^3
\frac{\partial^3\tilde{h}} {\partial {\tilde{x}}^3} ) +
\sqrt{2\sigma} \frac{\partial}{\partial \tilde{x}} \left
[{\tilde{h}}^{3/2} \xi (\tilde{x},\tilde{t}) \right ] \ ,
\label{lubrication31dnon}
\end{equation}
where $\tilde{h} = h/h_{0} \ , \  \tilde{x} = x/h_{0} \ , \
\tilde{t}=t/t_0$, where $h_0$ is the maximal height of the initial
drop, $t_0 =  3 \eta h_0 / \gamma $, and $\sigma = k_B T / \gamma
W h_0$. Our initial condition is a one-dimensional droplet with
circular cross section, and a substrate wet by a precursor film of
height $b$ to ensure complete wetting (see Fig. 1). Physically,
$b$ represents a scale below which the surface dynamics is not
governed by surface tension. For molecular liquids, this scale is
related to the dominance of VdW forces near solid walls, whereas
for complex fluids it is determined by the size of the fluid
constituents $a$ (e.g. for colloidal suspensions $a$ is the
colloid diameter). The results of our simulations are independent
of $b$, as long as $b$ is small enough compared to other length
scales involved in the dynamics (\ref{lubrication3}).

Typical drop shapes obtained by averaging
$\tilde{h}(\tilde{x},\tilde{t})$ over many realizations of Eq.
(\ref{lubrication31dnon}) with several values of $\sigma$ are
presented in Fig. 2, and compared to the dynamics of a spreading
drop governed by the deterministic Eq. (\ref{lubrication1}).
\begin{figure}
\begin{center}
\includegraphics[width=1.6in,clip=]{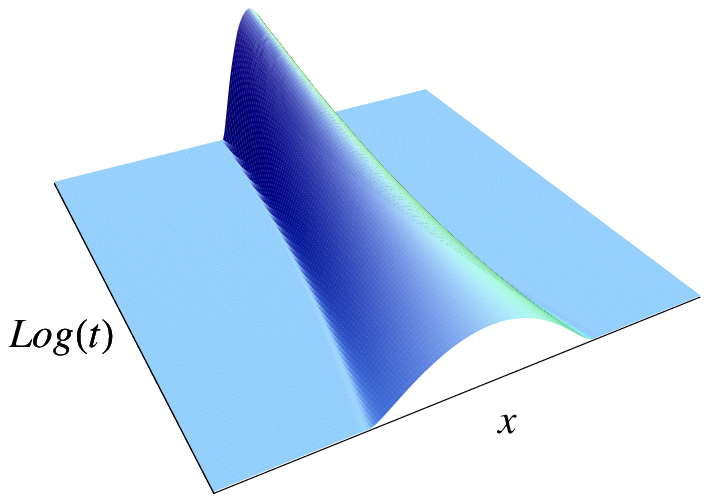}
\includegraphics[width=1.6in,clip=]{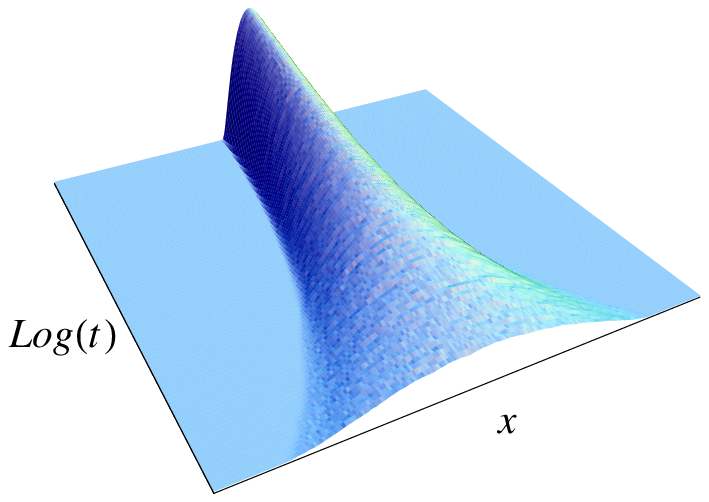}
\caption{\label{figshape} Average shape of the droplet $\langle
\tilde{h}(\tilde{x},\tilde{t})\rangle$ over 50 realizations as a
function of time for zero temperature (left) and $\sigma =
10^{-2}$ (right). The right picture clearly shows the assisted
spreading of the droplet in presence of thermal fluctuations.}
\end{center}
\end{figure}
Fig. 2 clearly indicates the effect of thermal fluctuations in
enhancing the rate of spreading drops.

To gain some quantitative understanding of this effect, we
measured the average rate by which characteristic lateral scales
$\tilde{\ell}$ of the drop evolve for various magnitudes of the
stochastic force, and compared them to Tanner's law in this
geometry: ${\tilde{\ell}}_{det} \sim {\tilde{t}}^{1/7}$. To
estimate ${\tilde{\ell}}$ we have used the averaged second moment
of the height profile,
\begin{equation}\label{radiusmeasure}
{\tilde{\ell}} = \left\langle\left[\frac{1}{V}\int d\tilde{x}
(\tilde{x}-\overline{\tilde{x}})^2 \tilde{h}(\tilde{x},\tilde{t})
\right]^{1/2}\right\rangle
\end{equation}
where $\overline{\tilde{x}} = [\int d\tilde{x} \ \tilde{x}
\tilde{h}(\tilde{x},\tilde{t})]/V$ is the instantaneous droplet
center position, $V = \int d\tilde{x} \
\tilde{h}(\tilde{x},\tilde{t})$ is the constant volume of the
droplet, and $\langle\cdots\rangle$ represents an average over
realizations of the noise $\xi(\tilde{x},\tilde{t})$. The results
of this analysis are shown in Fig. 3a, from which we extract a
modified asymptotic spreading rate in volume-preserving
one-dimensional flow geometry: ${\tilde{\ell}}_{stoch} \sim
{\tilde{t}}^{1/4}$. Obviously, the larger $\sigma$ is the earlier
is the deviation from Tanner's to the fluctuations-dominated
asymptotic rate of spreading.
\begin{figure}
\begin{center}
\includegraphics[width=3.2in,clip=]{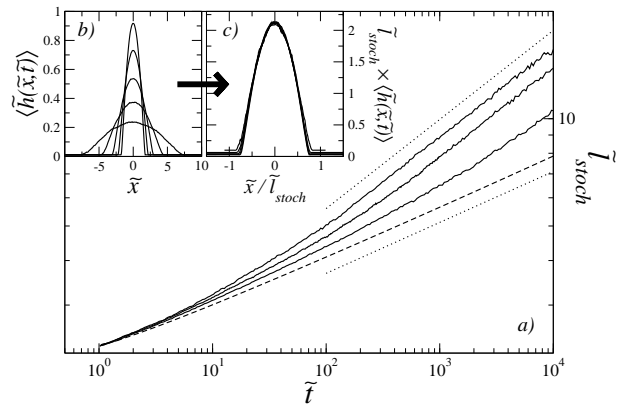}
\caption{\label{figradius} Results of volume conserving dynamics
in one-dimensional geometry. a) Lateral scale of the droplet
(\ref{radiusmeasure}) as a function of time. Solid lines represent
averages over 50 realizations of (\ref{lubrication31dnon}) with
$\sigma= 10^{-2}, 5\times 10^{-3}, 10^{-3}$ (from top to bottom),
while the dashed line is the noiseless ($\sigma = 0$) dynamics.
Dotted lines correspond to the power laws ${\tilde{t}}^{1/4}$ and
${\tilde{t}}^{1/7}$. b) Averaged profile of the droplet for times
${\tilde{t}}=10^{3n/4}$, $n=0,1,2,3,4$. c) Rescaled droplet
profiles. The parameters of the simulation are $\Delta {\tilde{t}}
= 0.01$, $\Delta {\tilde{x}} = 0.05$, $\tilde{b} = 0.01$. }
\end{center}
\end{figure}

The appearance of a new power law for spreading suggests that the
dynamics of the spreading drop is self-similar. The
self-similarity is demonstrated through the excellent data
collapse shown in Figs. 3b, 3c. Indeed, the power law
$\tilde{\ell}_{stoch} \sim {\tilde{t}}^{1/4} $ follows directly by
assuming a self-similar solution (in the statistical sense) of the
form (\ref{scalingform}) to Eq. (\ref{lubrication31dnon})
\cite{barabasi}. Conservation of drop volume in one-dimensional
geometry implies $\beta=1$ in the self-similar form
(\ref{scalingform}). Substituting this in Eq.
(\ref{lubrication31dnon}) gives rise to two possible scalings. The
first, obtained by assuming that the surface tension term is
dominant on the RHS of Eq. (\ref{lubrication31dnon}), is just
Tanner's law $\tilde{\ell}_{det} \sim {\tilde{t}}^{1/7} $. By
contrast, if the stochastic force is dominant we consider a
self-similar dynamics of the average $\langle \tilde{h} \rangle =
f(|\tilde{x}|/{\tilde{t}}^{z}) / |\tilde{x}|$. On the RHS of Eq.
(\ref{lubrication31dnon}) we substitute the average stochastic
force $\langle \xi \rangle = 1 / \sqrt{\tilde{t} |\tilde{x}|}$
over a time interval ${\tilde{t}}$ and space interval
${|\tilde{x}|}$. Thus we obtain $\tilde{\ell}_{stoch} \sim
{\tilde{t}}^{1/4} $, in agreement with the asymptotic behavior of
our simulation. Assuming the scaling relations $\tilde{h} \sim
|\tilde{x}|^{-1}$ and $\tilde{t} \sim |\tilde{x}|^{4}$ we evaluate
the surface tension term in Eq. (\ref{lubrication31dnon}) as
$|\tilde{x}|^{-8}$ and the average of the stochastic force as
$\sqrt{2 \sigma} |\tilde{x}|^{-5}$ . Requiring dominance of the
stochastic term and returning to dimensional variables, we obtain
that the stochastic scaling behavior is expected in the regime
\begin{equation}
|x| \gg x^{*} \ \ , \ \ h \ll h_0^2 / x^{*} \ , \label{condition}
\end{equation}
where $x^{*} = h_0^{7/6} W^{1/6} /{{\ell}_T}^{1/3} $ and ${\ell}_T
= \sqrt{k_B T /\gamma}$. Thus, a necessary condition for
observation of stochastic scaling behavior is $h_0 \gg W \gg
{\ell}_T$. Typical values of ${\ell}_T$ are few angstroms for
molecular fluids (far from the critical point), or colloid size
for colloidal suspensions \cite{Lekerkerker}. Applying similar
analysis for the radially symmetric spreading, Eq.
(\ref{lubrication32d}), shows that in this case self-similarity
dominated by the stochastic force gives rise to an enhanced rate
of spreading ${\tilde{\ell}}_{stoch} \sim {\tilde{t}}^{1/6}$,
compared to Tanner's law ${\tilde{\ell}}_{det} \sim
{\tilde{t}}^{1/10}$. The average stochastic force is dominant over
surface tension if
\begin{equation}
r \gg h_0^{4/3} /{{\ell}_T}^{1/3} \ , \ h \ll h_0^{2/3}
{{\ell}_T}^{1/3} \ , \label{condition2d}
\end{equation}
and thus requires $h_0 \gg {\ell}_T$.

For molecular fluids, the enhanced rates associated with
stochastic scaling behavior can be observed if the stochastic
force in Eq. (\ref{lubrication31dnon}) is dominant not only with
respect to the surface tension term as is expressed in Eqs.
(\ref{condition},\ref{condition2d}), but also with respect to the
VdW force that can be quite strong for thin films
\cite{Israelachvili}. In the nondimensional units of Eq.
(\ref{lubrication31dnon}) this force takes the form
$\frac{A}{\gamma h_0^2}  \frac{\partial}{\partial \tilde{x}} (
\tilde{h}^{-1} \frac{\partial \tilde{h}}{\partial{\tilde{x}}} )$.
Following a similar approach to the one that led to Eq.
(\ref{condition}) we use for one-dimensional geometry the scaling
relations $\tilde{h} \sim |\tilde{x}|^{-1}$ and $\tilde{t} \sim
|\tilde{x}|^{4}$ to compare between the average stochastic and VdW
forces and obtain the additional condition for the stochastic
scaling regime
\begin{equation} x \ll h_0^{3/2} {{\ell}_T}^{1/3} /
{{\ell}_{VdW}}^{2/3}  W^{1/6} \ ,  \label{condition2}
\end{equation}
where ${\ell}_{VdW} = \sqrt{A / \gamma}$. An overlap between the
intervals in Eq. (\ref{condition}) and (\ref{condition2}) is
achieved if $h_0 \ll  W ({{\ell}_{VdW}} / {{\ell}_T}) ^{2} $.
Typical values of $A$ are $100 k_BT$ \cite{Israelachvili} and thus
${{\ell}_{VdW}} > {{\ell}_{T}}$, and this overlap can be obtained
for $ h_0 \gg W \gg {\ell}_T$. Similar analysis for
two-dimensional geometry yields the result $r \ll (h_0^{11}
{{\ell}_{T}}^2 / {{\ell}_{VdW}}^4)^{1/9}$. Consistency of this
condition with Eq. (\ref{condition2}) is possible only if $h_0
\ll{{\ell}_{T}}^5 / {{\ell}_{VdW}}^4$, which seems unfeasible for
typical fluids. We conclude that volume preserving
fluctuations-dominated spreading can be observed for molecular
fluids in one-dimensional flows if the initial height of the drop
is large enough.

By contrast, for complex fluids VdW forces with the solid plate
are not expected to significantly affect the spreading process,
and thus we require $x \ (r)  \ll h_0^2 / a \ , h \gg a$ for one-
and two-dimensional flows, respectively. An overlap with Eqs.
(\ref{condition},\ref{condition2d}) is achieved if $h_0 \gg (a^6 W
/ {{\ell}_T}^2)^{1/5} \ , \  a^{3/2} / {{\ell}_T}^{1/2}$ for one-
and two-dimensional flow geometries, respectively. Both conditions
are easily realized if the initial drop is large enough.

Another spreading dynamics in which a stochastic scaling behavior
might be observed is a "leaking'' process, in which the height of
the film at $x=0$ is fixed to a constant value $h_0$ by a
continuous supply of fluid. A self-similar dynamics in this case
has the form (\ref{scalingform}) with $\beta=0$. Following similar
analysis we obtain for one-dimensional geometry the scaling
behaviors: ${\tilde{\ell}}_{det} \sim {\tilde{t}}^{1/4}$,
${\tilde{\ell}}_{stoch} \sim {\tilde{t}}^{1/3}$, while for
two-dimensional geometry both ${\tilde{\ell}}_{det}$ and
${\tilde{\ell}}_{stoch} \sim {\tilde{t}}^{1/4}$. For the
one-dimensional geometry we obtain an increased asymptotic rate of
spreading due to thermal fluctuations. To check the realizability
of the stochastic scaling regime we compare the average stochastic
force with surface tension and VdW terms, using the scaling
relations: $\tilde{h} \sim const$ and $\tilde{t} \sim
{\tilde{x}}^{3}$. For molecular fluids where VdW forces are
important the stochastic scaling behavior is expected in the
regime $ h_0^{3/2} W^{1/2} / {\ell}_T \ll x \ll h_0^{5/2} {\ell}_T
/ W^{1/2} {{\ell}_{VdW}}^2$, which is again possible provided $h_0
\ll  W ({{\ell}_{VdW}} / {{\ell}_T}) ^{2} $. Stochastic scaling
behavior of a complex fluid drop under this condition is achieved
if $h_0 \gg a$ and $x \gg h_0^{3/2} W^{1/2} / {\ell}_T$.

To conclude, we derived a Langevin lubrication equation
(\ref{lubrication3}), and showed that it gives rise asymptotically
to significant deviations from Tanner's law of spreading. By
comparing the average stochastic force with classical forces such
as surface tension and Van der Waals, we showed that
fluctuation-assisted spreading is expected asymptotically in
various flow geometries of molecular and complex fluids. Complex
fluids are attractive candidates for studying this phenomenon,
since confocal microscopy techniques enable direct observation of
macroscopic thermal effects in systems such as colloidal
suspensions. Direct imaging of thermal capillary waves of an
interface between two colloidal liquids was recently achieved by
such methods \cite{Lekerkerker}, and we believe that similar tools
may enable observation of the enhanced rate of spreading predicted
in this letter. We hope that this result, together with other
examples, will motivate further studies of the role of thermal
fluctuations in small dimensional fluid systems.

We thank M. Brenner, R. Cuerno and M. Kardar for useful
discussions. This work was supported by Harvard MRSEC
(DMR-0213085) (B.D and H.A.S) and by Ministerio de Ciencia y
Tecnolog\'{\i}a (Spain) through project FIS2004-01001 and
BFM2002-04474-C02 and the Real Colegio Cumplutense at Harvard
University (E.M).


\begin{thebibliography}{99}

\bibitem{Tanner} L.\ Tanner, J.\ Phys.\ D {\bf 12}, 1473 (1979).

\bibitem{McKinley} P.\ Kavehpour, B.\ Ovryn, and G.\ H.\ McKinley,
Phys. Rev. Lett. {\bf 91}, 196104 (2003).

\bibitem{DeGennes} P.G. DeGennes, Rev. Mod. Phys. {\bf 57}, 827
(1985).

\bibitem{Koplik-Freund-He} J. Koplik, J.R. Banavar and J.F. Willemsen,
Phys. Fluids. A. {\bf 1}, 781 (1989) ; J.B. Freund, Phys. Fluids.
{\bf 15}, L-33 (2003) ; G. He and N.G. Hadjiconstantinou, J.
Fluid. Mech. {\bf 497}, 123 (2003).

\bibitem{Esteban} D.\ B.\ Abraham, R.\ Cuerno and E.\ Moro,
Phys. Rev. Lett. {\bf 88}, 206101 (2002).

\bibitem{Landman} M.\ Moseler and U.\ Landman, Science {\bf 289}, 1166
(2000).

\bibitem{Pismen} L.M. Pismen and Y. Pomeau,  Phys. Rev. E {\bf 62}, 2480 (2000).

\bibitem{Batchelor} G.K. Batchelor, {\em An Introduction to Fluid Dynamics},
Cambridge University Press (2000).

\bibitem{Landau} E.M. Lifshitz and L.P. Pitaevskii,{\em Statistical Physics
II}, Butterworth-Heinemann (1980).

\bibitem{Siggia} A.M.S. Tremblay, M. Arai and E.D. Siggia, Phys. Rev. A {\bf 23}, 1451
(1981).

\bibitem{numerical} See J.\ A.\ Diez, L.\ Kondic, and A.\ Bertozzi
Phys. Rev. E {\bf 63}, 011208 (2000) and references therein.

\bibitem{barabasi} A.-L.\ Barab\'asi and H.\ E.\ Stanley, {\em
Fractal Concepts in Surface Growth}, Cambridge University Press
(1995).

\bibitem{Israelachvili} J.N. Israelachvili, {\em Intermolecular and
Surface Forces}, Academic Press (1992).

\bibitem{Lekerkerker}
D.G.A.L. Aarts, M. Schmidt, and H.N.W. Lekkerkerker, Science {\bf
304}, 847 (2004).

\end{thebibliography}
\end{document}